\documentclass[a4paper,11pt]{article}

\usepackage{amstex}
\usepackage[dvips]{graphicx}

\title{A metaphor for adiabatic evolution to symmetry}

\author{
  R.J.A.G.~Huveneers\thanks{Supported by NWO-grant nr. 611-306-537}
  \hspace{2cm} F.~Verhulst \\
  \\
  Department of Mathematics \\
  University of Utrecht \\
  P.O. Box 80.010, 3508 TA Utrecht \\
  The Netherlands
}

\begin{document}

\maketitle

\begin{abstract}
In this paper we study a Hamiltonian system with a spatially asymmetric
potential. We are interested in the effects on the dynamics when the potential
becomes symmetric slowly in time. We focus on a highly simplified non-trivial
model problem (a metaphor) to be able to pursue explicit calculations as far as
possible. Using the techniques of averaging and adiabatic invariants, we are
able to study all bounded solutions, which reveals significant asymmetric
dynamics even when the asymmetric contributions to the potential have become
negligibly small.
\end{abstract}

\section{Introduction}

Many physical objects exhibit some form of symmetry. Most galaxies for
instance, have axes or planes of symmetry. The motivation for this study is
that a symmetric equilibrium configuration generally is the outcome of the
evolution from an asymmetric state. We would like to trace the effect of the
asymmetries.

A problem is that studies of the evolution of actual physical systems are
difficult and so relatively rare. We propose therefore to ignore, at least for
the time being, the actual physical mechanisms and to consider systems
described by a Hamiltonian of the form

\begin{equation}
  \mathcal{H}(p, q, \epsilon t) = \mathcal{H}_s(p, q) + a(\epsilon t)
  \mathcal{H}_a(p, q)
\end{equation}

where $\mathcal{H}_s$ is the part of the Hamiltonian which is symmetric in some
sense; $\mathcal{H}_a$ is the asymmetric
part which is slowly vanishing as we put

\begin{equation}
  a(0) = 1, \lim_{t \rightarrow \infty} a(\epsilon t) = 0, 0 < \epsilon \ll 1
\end{equation}

To study the dynamics induced by the Hamiltonian $\mathcal{H}(p, q, \epsilon
t)$ is still a formidable problem. So we
simplify as much as possible to obtain

\begin{equation}
  \label{Sys:x}
  \left\{
    \begin{array}{rcl}
      \dot{x}_1 & = & x_2 \\
      \dot{x}_2 & = & -x_1 + a(\epsilon t) x_1^2
    \end{array}
  \right.
\end{equation}

which is derived from the one degree of freedom Hamiltonian

\begin{equation}
  \mathcal{H}(p, q, \epsilon t) = \frac{1}{2} (p^2 + q^2) + \frac{1}{3}
  a(\epsilon t) q^3
\end{equation}

identifying $p = x_2$, $q = x_1$. We shall associate with system (\ref{Sys:x})
the ``unperturbed'' system which
arises for $\epsilon = 0$

\begin{equation}
  \label{Sys:x:eps=0}
  \left\{
    \begin{array}{rcl}
      \dot{x}_1 & = & x_2 \\
      \dot{x}_2 & = & -x_1 + x_1^2
    \end{array}
  \right.
\end{equation}

\begin{figure}
  \begin{center}
    \includegraphics[height=8cm]{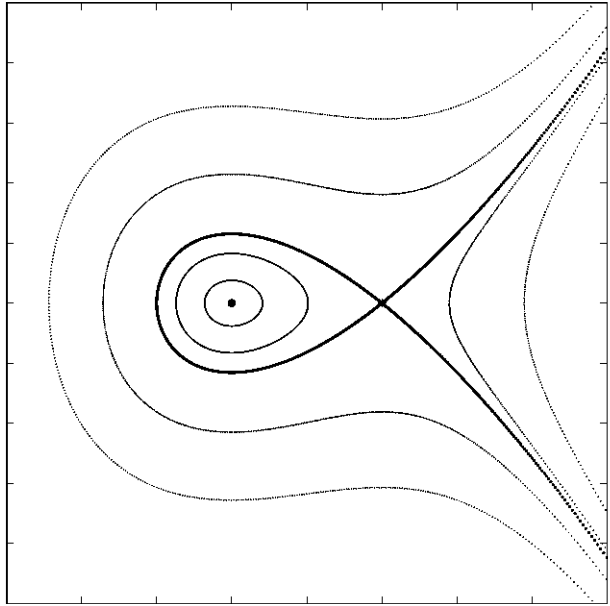}
  \end{center}
  \caption{The dynamics of the unperturbed system (\ref{Sys:x:eps=0})}
  \label{Fig:unperturbed system}
\end{figure}

We note that in the autonomous system (\ref{Sys:x:eps=0}) there are basically
two regions (figure \ref{Fig:unperturbed system}): within the homoclinic
solution the orbits are bounded, outside the homoclinic solution the orbits
diverge to infinity (with the exception of the stable manifold and the saddle
point itself). In system (\ref{Sys:x}) we have no fixed saddle point, still it
turns out that we have two
separate regions of initial values in which the orbits are bounded or diverge
to infinity.

Since the dynamics of systems (\ref{Sys:x}) and (\ref{Sys:x:eps=0}) are the
same on an $O(1)$ timescale, it is instructive (though slightly wrong) to view
system (\ref{Sys:x}) as having a saddle point moving slowly towards infinity
and having a slowly expanding homoclinic orbit. Within this picture, an orbit
can remain bounded in two ways, either by starting inside the homoclinic orbit,
or by getting ``captured'' by the slowly expanding homoclinic orbit, which can
only happen if the orbit starts sufficiently close to the stable manifold of
the saddle point.

Using a special transformation we shall discuss the boundaries of these regions
in section \ref{Section: The boundary of the stable part of phase space}.
A special case, $a(\epsilon t) = \exp(-\epsilon t)$ can be studied easily and
help us to understand the general case.

In section \ref{Section: Averaging inside the stable region} we perform
averaging in the so-called stable region where bounded solutions are found.
This involves the use of elliptic and hypergeometric functions, rather hard
analysis, where we are
supported by Mathematica 2.2 running under SunOS 4.1.3. \\
After determining the validity of the averaged equation we establish the
existence of an adiabatic invariant in the stable region, valid for all time.
Even more remarkable is that explicit calculations of this invariant show that
the evolution of
phase points will show  significant traces of its asymmetrical past for all
time.

In section \ref{Section: The boundary layer} we need subtle reasoning to
discuss what is going on in the boundary layer near
the boundary of the stable domain.

The analysis in this paper is based on averaging methods but, because of its
direct relation to dissipative mechanics
(section \ref{Section: The boundary of the stable part of phase space}), it
clearly profits from the results by
Haberman and Ho~\cite{H&H 1,H&H 2} and Bourland and Haberman~\cite{B&H}. At the
same time our analysis should be placed within the theory of adiabatic
invariants, which has been summarized recently in an admirable survey by
Henrard~\cite{H}.

We finally note that in the context of galactic dynamics, some rather different
examples based on classical results of the theory of adiabatic invariants were
given by Binney and Tremaine~\cite{B&T}.

\section{The boundary of the stable part of phase space}
\label{Section: The boundary of the stable part of phase space}

As we explained in the introduction, the phase space of system (\ref{Sys:x})
can be separated into two parts. Since we are dealing with a time-dependent
system, we must specify the time for which a particular separation holds. We
use the following definition:

The \emph{stable} part of phase space consists of the points $(x_1, x_2)$, for
which the orbit $\gamma (x_1, x_2, 0)$ starting in $(x_1, x_2)$ at $t=0$
remains bounded for $t$ going to infinity. All other points define the
\emph{unstable} part of phase space.

Clearly, a point $(x_1, x_2)$ can only be contained in the stable region if it
lies within an $O(\epsilon )$ neighbourhood of the area bounded by the
homoclinic orbit of system (\ref{Sys:x:eps=0}). If this is not the case,
$\gamma (x_1, x_2, 0)$ will reach the upper branch of the unstable manifold of
the saddle point of system (\ref{Sys:x:eps=0}) in a finite time and clear off
to infinity.
We must not overlook the orbits starting close to the lower branch of the
stable manifold of the saddle point of system (\ref{Sys:x:eps=0}), which can
reach the just described $O(\epsilon )$ neighbourhood too. It will turn out
that although this region may look small, it produces the major part of the
stable region.

These considerations help us locating the boundary of the stable region
approximately.
The location of the boundary of the stable region separates the part of phase
space in which all orbits diverge to infinity ($(x_1, x_2) \rightarrow{}
(+\infty , +\infty )$) from the part of phase space in which the orbits tend to
circle around the origin for $t$ going to infinity, so if we expect to see
effects of the vanishing of the asymmetric potential somewhere, it is just
within this boundary.

The key step in analyzing system (\ref{Sys:x}) is performing the transformation

\begin{equation}
  \label{Trafo:x->y}
  y_1 = a(\epsilon t) x_1
\end{equation}

The idea behind this transformation is to try to fix the ``slowly moving saddle
point'' of system (\ref{Sys:x}).
Demanding that $\dot{y}_1 = y_2$, we arrive at the system

\begin{equation}
  \label{Sys:y}
  \left\{
    \begin{array}{rcl}
      \dot{y}_1 & = & y_2 \\
      \dot{y}_2 & = & -y_1 + y_1^2 + 2 \epsilon \frac{a'(\epsilon
      t)}{a(\epsilon t)} y_2 + {\epsilon}^2 \left( \frac{a''(\epsilon
      t)}{a(\epsilon t)} - 2 \frac{{a'(\epsilon t)}^2}{{a(\epsilon t)}^2}
      \right) y_1
    \end{array}
  \right.
\end{equation}

where $a'(\epsilon t)$ stands for ${\left. \frac{\mathrm{d}a(\xi
)}{\mathrm{d}\xi } \right\vert}_{\xi{} = \epsilon t}$ and similarly for
$a''(\epsilon t)$.

By transformation (\ref{Trafo:x->y}) the slow time-dependence has moved to
$O(\epsilon)$ terms; still, system (\ref{Sys:y}) looks more complicated than
system (\ref{Sys:x}).
However, we will be able to neglect the $O({\epsilon}^2)$ term in most of our
calculations.
We should also note that system (\ref{Sys:y}) is \emph{not} Hamiltonian
anymore, since we have applied a non-canonical transformation. Indeed the
$O(\epsilon )$ term is a friction term, causing the origin $(y_1, y_2) = (0,
0)$ to become an attracting focus instead of a center.
In the analysis of system (\ref{Sys:y}) we start with a special choice of
$a(\epsilon t)$.

\subsection{The special case $a(\epsilon t) = e^{-\epsilon t}$}

We will first calculate the location of the boundary of the stable region for
the special, but physically important case

\begin{equation}
  \label{Special:e-macht}
  a(\epsilon t) = e^{-\epsilon t}
\end{equation}

We will show later that the general case does not differ much from this special
case.
With the choice (\ref{Special:e-macht}) for $a(\epsilon t)$, system
(\ref{Sys:y}) reduces to

\begin{equation}
  \label{Sys:y:e-macht}
  \left\{
    \begin{array}{rcl}
      \dot{y}_1 & = & y_2 \\
      \dot{y}_2 & = & -y_1 + y_1^2 - 2 \epsilon y_2 - {\epsilon}^2 y_1
    \end{array}
  \right.
\end{equation}

It is remarkable that for this special yet interesting choice of $a(\epsilon
t)$, our system becomes autonomous, which reduces the calculations because the
dependence on the initial time has vanished into the transformation
(\ref{Trafo:x->y}). \\
We also note that we have succeeded in fixing the saddle point: The saddle
point of system (\ref{Sys:y:e-macht}) is located in $(1 + {\epsilon}^2, 0)$.

The saddle point not being located in $(1,0)$ as we intended would introduce a
lot of extra small terms in our calculations. To avoid these we map the saddle
point onto $(1,0)$ by substituting $y_i \rightarrow{} (1 + {\epsilon}^2) y_i,
i= 1, 2$, to obtain

\begin{equation}
  \label{Sys:y:e-macht:mapped}
  \left\{
    \begin{array}{rcl}
      \dot{y}_1 & = & y_2 \\
      \dot{y}_2 & = & -(1 + {\epsilon}^2) (y_1 - y_1^2) - 2 \epsilon y_2
    \end{array}
  \right.
\end{equation}

So we have reduced the calculation of the boundary of the stable region of
system (\ref{Sys:x}) to the calculation of the (time-independent) region of
attraction of system (\ref{Sys:y:e-macht:mapped}).

\begin{figure}
  \begin{center}
    \includegraphics[height=8cm]{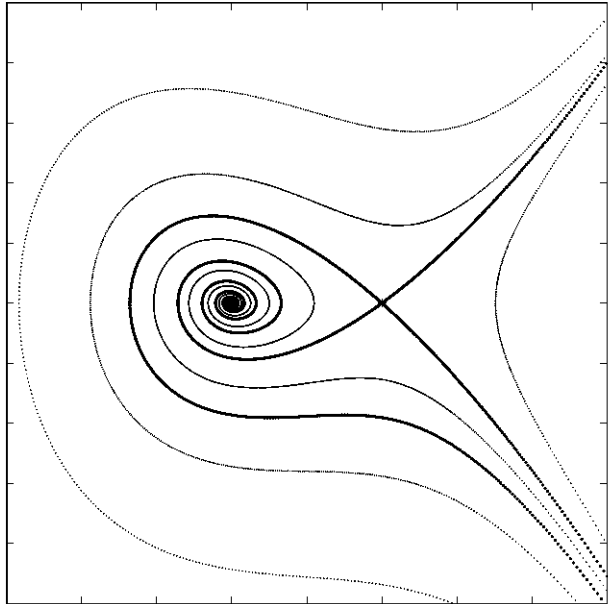}
  \end{center}
  \caption{The stable and unstable manifold of system
  (\ref{Sys:y:e-macht:mapped}) with $\epsilon = 0.1$}
  \label{Fig:y:e-macht:manif}
\end{figure}

It is easily seen (figure \ref{Fig:y:e-macht:manif}) that the region of
attraction of system (\ref{Sys:y:e-macht:mapped}) is bounded by the stable
manifold of the saddle point.

It is well known that generally the stable manifold of a perturbed system (with
parameter $\epsilon$) lies in an $O(\epsilon )$ neighbourhood of the stable
manifold of the unperturbed system (with $\epsilon = 0$).
The unperturbed system

\begin{equation}
  \label{Sys:y:e-macht:mapped:eps=0}
  \left\{
    \begin{array}{rcl}
      \dot{y}_1 & = & y_2 \\
      \dot{y}_2 & = & -y_1 + y_1^2
    \end{array}
  \right.
\end{equation}

is simple and totally understood. It has a first integral
$E(\epsilon = 0)$ where

\begin{equation}
  \label{Def:integral:eps=0}
  E(\epsilon = 0) = \frac{1}{2} y_2^2 + \frac{1}{2} y_1^2 - \frac{1}{3} y_1^3
\end{equation}

and the unstable manifold coincides with the homoclinic orbit $E(\epsilon = 0)
= \frac{1}{6}$.

Using $E(\epsilon = 0)$ in our calculations for the perturbed system would
introduce some higher order terms. Instead, we extend the
definition of $E$ with suitable $O(\epsilon^2)$ terms which cancel these
terms. Again, this is only for calculational convenience.

\begin{equation}
  \label{Def:integral:eps}
  E = \frac{1}{2} y_2^2 + \frac{1}{2} (1 + {\epsilon}^2) y_1^2 - \frac{1}{3} (1
  + {\epsilon}^2) y_1^3
\end{equation}

\begin{figure}
  \begin{center}
    \includegraphics[height=8cm]{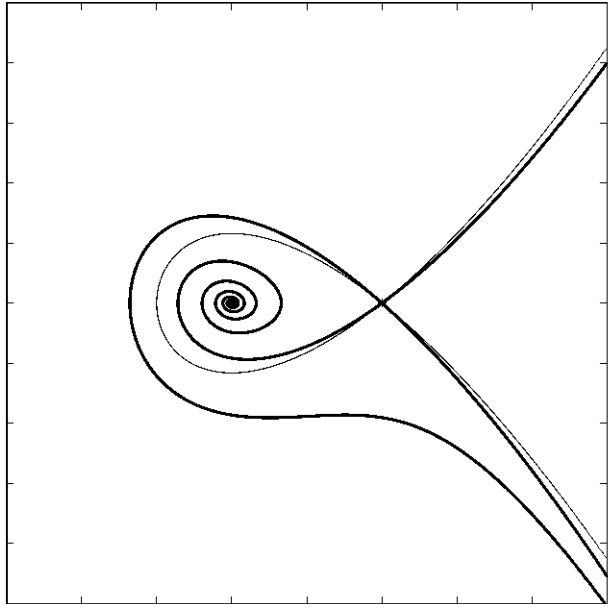}
  \end{center}
  \caption{The homoclinic orbit (represented by the thin line) of system
  (\ref{Sys:y:e-macht:mapped:eps=0}) added to figure \ref{Fig:y:e-macht:manif}}
  \label{Fig:y:e-macht:manif+homocl}
\end{figure}

It is instructive to combine figure \ref{Fig:y:e-macht:manif} with the
homoclinic orbit of the unperturbed system (\ref{Sys:y:e-macht:mapped:eps=0}),
which produces figure \ref{Fig:y:e-macht:manif+homocl}.

We will now approximate the location of the stable manifold of the saddle point
of system (\ref{Sys:y:e-macht:mapped}) by calculating the variation of $E$
along the stable manifold. Since this variation is an $O(\epsilon )$ effect, we
may use the unperturbed stable manifold in this calculation, which involves
elliptic functions. From this variation of $E$ along the stable manifold, we
can deduce the location of the perturbed stable manifold.

If we follow the flow along the stable manifold from a point $(y_{10}, y_{20})$
to a point $(y_{11}, y_{21})$ we get:

\begin{equation}
  \frac{\mathrm{d}E}{\mathrm{d}t} = -2 \epsilon y_2^2
  \Rightarrow
  \Delta E = \int{-2 \epsilon y_2^2 dt} = -2 \epsilon
  \int_{y_{10}}^{y_{11}}{y_2 dy_1}
\end{equation}

The integral appearing in this expression has to be calculated with
$O(\epsilon)$ precision, which allows us to substitute the explicitly known
orbits of the unperturbed system (\ref{Sys:y:e-macht:mapped:eps=0}).
These orbits ($y_2(y_1)$) are readily obtained from the definition of the
first integral (\ref{Def:integral:eps=0}).

To calculate the variation of $E$ along the upper branch of the stable
manifold,
we take $(y_{11}, y_{21})$ = $(1,0)$ and get, after some analysis as indicated
above:

\begin{equation}
  E(y_1) = \frac{1}{6} (1 + {\epsilon}^2) + 2 \epsilon \left( \frac{3}{5} +
  \frac{\sqrt{3} (y_1 - 2) {(2 y_1 + 1)}^{(3/2)}}{15} \right) + O({\epsilon}^2
  {(y_1 - 1)}^2)
\end{equation}

which is valid for $-\frac{1}{2} \leq y_1 \leq 1$ and $y_2 > 0$.

For $y_1 < -\frac{1}{2}$ we have $\dot{E} = O({\epsilon}^2)$ and therefore we
get:

\begin{equation}
  E(y_1) = \frac{1}{6} + \frac{6}{5} \epsilon + O({\epsilon}^2)
\end{equation}

which is valid for $y_1 < -\frac{1}{2}$

Taking $(y_{11}, y_{21})$ = $(-\frac{1}{2}, y_{21})$ we get:

\begin{equation}
  \label{Eq:log:epsilon:error}
  E(y_1) = \frac{1}{6} + 2 \epsilon \left( \frac{3}{5} - \frac{\sqrt{3} (y_1 -
  2) {(2 y_1 + 1)}^{(3/2)}}{15} \right) + O({\epsilon}^2 \log \epsilon)
\end{equation}

which is valid for $-\frac{1}{2} \leq y_1 \leq 1$ and $y_2 < 0$.
The special form of the error term arises from the fact that the homoclinic
orbit is only an $O(\sqrt{\epsilon})$ approximation of the stable manifold
for $y_1$ close to 1 and negative $y_2$ (just under the saddle point).
This follows from the analysis in Haberman and Ho~\cite{H&H 1}.


Taking $(y_{11}, y_{21})$ = $(1, y_{21})$ we get

\begin{equation}
  E(y_1) = \frac{1}{6} + 2 \epsilon \left( \frac{9}{5} + \frac{\sqrt{3} (y_1 -
  2) {(2 y_1 + 1)}^{(3/2)}}{15} \right) + O({\epsilon}^2 \log \epsilon)
\end{equation}

which is valid for $y_1 > 1$ and $y_2 < 0$.

To calculate the variation of $E$ along the lower branch of the stable
manifold,
we take $(y_{11}, y_{21})$ = $(1,0)$ and making use of the expressions for the
explicitly known lower branch of the stable manifold of the unperturbed system
(\ref{Sys:y:e-macht:mapped:eps=0}) we find

\begin{equation}
  E(y_1) = \frac{1}{6} (1 + {\epsilon}^2) + 2 \epsilon \left( \frac{3}{5} +
  \frac{\sqrt{3} (y_1 - 2) {(2 y_1 + 1)}^{(3/2)}}{15} \right) + O({\epsilon}^2
  {(y_1 - 1)}^2)
\end{equation}

which is valid for $y_1 > 1$ and $y_2 < 0$.

So, we have now calculated the variation of $E$ all over the stable manifold of
the saddle point of system (\ref{Sys:y:e-macht:mapped}). What is left to do is
to deduce the location of the stable manifold itself from this variation, which
is not very hard.

Given a value of $y_1$, one first calculates the corresponding value of $E$
using the appropriate formula given above. Using the definition of $E$
(\ref{Def:integral:eps}), one calculates the corresponding value of $y_2$. This
amounts to solving a third order polynomial, which can even be done
explicitly.

In particular one can compute the intersection of the stable manifold with the
$y_1$-axis, which occurs (approximately) in $(-\frac{1}{2} - \frac{8}{5}
\epsilon , 0)$.

\subsection{The boundary of the stable region for arbitrary $a(\epsilon t)$}

We now return to the discussion of the general system (\ref{Sys:y}).
It turns out that the analysis is essentially the same as for the
special case $a(\epsilon t) = e^{-\epsilon t}$

We claim that the behaviour of system (\ref{Sys:y}) is (with a certain error)
described by the system

\begin{equation}
  \label{Sys:y:equiv}
  \left\{
    \begin{aligned}
      \dot{y}_1 &= y_2 \\
      \dot{y}_2 &= -y_1 + y_1^2 + 2 \epsilon \frac{a'(0)}{a(0)} y_2
    \end{aligned}
  \right.
\end{equation}

\emph{as far as the location of the boundary of the stable region is
concerned}.

The idea behind this statement is that if an orbit of system (\ref{Sys:y})
starts at a distance
$O(\delta )$ inside the \emph{unstable} region, it will reach the upper branch
of the unstable manifold after an interval of time $O(\log \delta )$, because
it has to pass
the saddle point at a distance $O(\delta )$ (sometimes twice). \\
Using Gronwall's inequality, it is easy to show that the orbit of system
(\ref{Sys:y:equiv}) starting at the same initial point, will diverge at most
$O(\frac{\epsilon^2 \log \delta}{\delta})$ from the exact orbit. \\
Since we know the boundary of the stable region with precision
$O({\epsilon}^2 \log \epsilon )$, we must take $\delta$ to be larger than
$O({\epsilon}^2 \log \epsilon )$ for our calculations to make sense. \\
Consequently, the orbit of system (\ref{Sys:y:equiv}) will diverge at most
$o(1)$ from the exact orbit and will thus diverge to infinity too. \\
Thus, a starting point $(y_1, y_2)$ lying more than $O({\epsilon}^2 \log
\epsilon )$
inside the unstable region produces an orbit diverging to infinity both in
system
(\ref{Sys:y}) and in system (\ref{Sys:y:equiv}).

We can apply exactly the same argument to the stable region, which proves that
the boundary of the stable region of system (\ref{Sys:y}) coincides with the
boundary of the stable region of system (\ref{Sys:y:equiv}) up to
$O({\epsilon}^2 \log \epsilon )$.

So we have the important conclusion that, to calculate the boundary of the
stable region of system (\ref{Sys:y}),
we can use the formulas derived for the special case
$a(\epsilon t) = e^{-\epsilon t}$ with $\epsilon$ replaced by
$-\frac{a'(0)}{a(0)} \epsilon$.

\section{Averaging inside the stable region for arbitrary $a(\epsilon t)$}
\label{Section: Averaging inside the stable region}

Knowing the location of the boundary of the stable region we proceed to study
the stable region itself (the unstable region is clearly not very
interesting). We have to do this study in two parts in which we consider the
interesting dynamics which takes place close to (we will make this more
precise) the boundary of the stable region (i.e. in the \emph{boundary layer})
and in the inner domain. At a safe distance from the
boundary layer, system (\ref{Sys:y}) will
behave more and more like a harmonic oscillator. The natural way to approach
such a problem is to apply the theory of averaging.

\subsection{Averaging in the inner domain}

Averaging in the vicinity of the origin is a simple exercise involving
averaging over harmonic functions. This is not what we have in mind; we shall
average over a part of the inner domain as large as possible. This involves
averaging over elliptic functions.

\subsubsection{Calculation of the averaged equation}

To perform averaging, we need one or more quantities with a small
($O(\epsilon)$) time derivative, i.e. which depend slowly on time. A
natural candidate for this quantity is the exact integral
(\ref{Def:integral:eps=0}) of the unperturbed system, for which we have

\begin{equation}
  \label{timeder:integral:eps=0}
  \frac{\mathrm{d}E}{\mathrm{d}t} = 2 \epsilon \frac{a'(\epsilon t)}{a(\epsilon
  t)} y_2^2 +
  \epsilon^2 \left( 1 - y_1 + \left( \frac{a''(\epsilon t)}{a(\epsilon t)} - 2
  \frac{{a'(\epsilon t)}^2}{{a(\epsilon t)}^2} \right)
\right)
y_1 y_2
\end{equation}

To be able to average this equation, we have to put restrictions on $a(\epsilon
t)$:

\begin{equation}
  \label{restr:average:a}
  \begin{aligned}
    \frac{a'(\xi)}{a(\xi)}  &\textrm{ is bounded for all positive } \xi \\
    \frac{a''(\xi)}{a(\xi)} &\textrm{ is bounded for all positive } \xi
  \end{aligned}
\end{equation}

Most decaying functions of interest satisfy these restrictions. Functions
decaying extremely
rapidly, such as $a(\xi) = \exp(-\exp(\xi))$, do not satisfy these
restrictions. But since $a(\xi)$ decays very rapidly, we can safely put
$a(\xi)$ equal to zero for all $\xi$ bigger than some $\xi_0$ for which
$a(\xi_0) \ll 1$, without affecting the dynamics. Other examples of
functions which do not satisfy (\ref{restr:average:a}) are
functions which vanish in a finite time like $a(\xi) = 1 - \xi$.
Again we can restrict the time span such that this poses no problem.

To average equation (\ref{timeder:integral:eps=0}), we consider $\tau =
\epsilon t$ as an independent variable and add the equation

\begin{equation}
  \dot{\tau} = \epsilon
\end{equation}

Since we only have to average the $O(\epsilon)$ part of equation
(\ref{timeder:integral:eps=0}), we have to average $y_2^2$ along a periodic
orbit of the unperturbed system (\ref{Sys:y:e-macht:mapped:eps=0}). This
amounts to calculating the integral of $y_2^2$ along the periodic orbit and
involves the period of the periodic orbit. This is in the spirit of
averaging as for instance presented in Sanders en Verhulst~\cite{S&V}.

To calculate $\int{y_2^2}dt$, we make use of $\dot{y}_1 = y_2$, which
reduces the calculation to the action $\int{y_2}dy_1$. The functional
dependence of
$y_2$ on $y_1$ for the unperturbed system can be retrieved from the exact
integral (\ref{Def:integral:eps=0}) and is the square root of a third order
polynomial. \\
Using this, we find that we also need this standard integral

\begin{equation}
  \int_{a}^{b}{\sqrt{(x-a)(b-x)(c-x)}} = \frac{1}{24} \sqrt{6} \pi {(b - a)}^2
\sqrt{c - a} \sideset{_2}{_1}F\left(-\frac{1}{2}, \frac{3}{2}, 3, \frac{b -
a}{c - a}\right)
\end{equation}

with $\sideset{_2}{_1}F$ the hypergeometric function, which holds when $a \leq
b \leq c$. \\
The $a$, $b$ and $c$ are the exacts roots of a third order polynomial and
are thus awkward expressions even for our simple unperturbed problem.
Surprisingly, the combinations $b-a$ and $c-a$ reduce to manageable
expressions:

\begin{equation}
  \label{b-a:c-a}
  \begin{aligned}
    b - a &= \sqrt{3} \sin\left(\frac{1}{3} \arcsin(12 E - 1) +
    \frac{\pi}{6}\right) \\
    c - a &= \sqrt{3} \cos\left(\frac{1}{3} \arcsin(12 E - 1)\right)
  \end{aligned}
\end{equation}

Substituting all this we get

\begin{equation}
  \label{integral:y2:over:y1}
  \begin{aligned}
    \int{y_2}dy_1 = &2 \frac{1}{24} \sqrt{6} \pi 3 \sin^2\left(\frac{1}{3}
    \arcsin(12 E - 1) + \frac{\pi}{6}\right) \times \\
    &\times \sqrt{\sqrt{3} \cos\left(\frac{1}{3} \arcsin(12 E - 1)\right)}
    \times \\
    &\times \sideset{_2}{_1}F\left(-\frac{1}{2}, \frac{3}{2}, 3,
    \frac{\sin\left(\frac{1}{3} \arcsin(12 E - 1) +
    \frac{\pi}{6}\right)}{\cos\left(\frac{1}{3} \arcsin(12 E -
    1)\right)}\right)
  \end{aligned}
\end{equation}

The factor 2 arises because we have to integrate once from $b$ to $a$ and once
from $a$ to
$b$.

To calculate the period of the periodic orbit of the unperturbed system, we
apply the standard technique of separation of variables to the exact integral
(\ref{Def:integral:eps=0}).
This leads us through a calculation similar to the one above, resulting in:

\begin{equation}
  \label{period:unperturbed:orbit}
  \begin{aligned}
    \mathrm{period} = &2 \sqrt{6} \frac{1}{\sqrt{\sqrt{3} \cos\left(\frac{1}{3}
    \arcsin(12 E - 1)\right)}} \times \\
    &\times \mathrm{K}\left(\frac{\sin\left(\frac{1}{3} \arcsin(12 E - 1) +
    \frac{\pi}{6}\right)}{\cos\left(\frac{1}{3} \arcsin(12 E -
    1)\right)}\right)
  \end{aligned}
\end{equation}

where K is the complete elliptic integral of the first kind.

We finally obtain the averaged equation by dividing equation
(\ref{integral:y2:over:y1}) by
equation (\ref{period:unperturbed:orbit}) and adding some extra factors from
equation (\ref{timeder:integral:eps=0}):

\begin{equation}
  \label{averaged:equation}
  \begin{aligned}
    \dot{\bar{E}} &= 2 \epsilon \frac{a'(\epsilon t)}{a(\epsilon t)}
      \frac{\int{y_2}dy_1}{\mathrm{period}} \\
    &= \epsilon \frac{a'(\epsilon t)}{a(\epsilon t)} A(\bar{E})
  \end{aligned}
\end{equation}

It does not add much to the understanding of the problem to write down the
averaged equation in
it's full form. That is why we omit this.
All that matters is that the right hand side is an explicitly known function
$A(\bar{E})$ of $\bar{E}$,
which we can approximate to arbitrary precision, and of time.

\subsubsection{Validity of the averaged equation}

Since the averaged equation is an approximation of the exact system
(\ref{Sys:y}), we have to
address the question of the accuracy of this approximation, on which timescale
it holds and
where in the stable region.

We expect that the closer we start to the homoclinic orbit of the unperturbed
system, the less accurate the averaged equation will become. The dynamics
splits up in two qualitatively different time intervals, the first in which the
orbit slowly separates from the homoclinic orbit and the second in which the
orbit
slowly spirals towards the attracting origin.

We start with the first time-interval. As we will show in section
\ref{subsection:Approaching the boundary layer}, apart from a sub-boundary
layer of size
$O(\exp(-\frac{1}{\epsilon}))$, this time-interval has a size of
$O(\frac{1}{\epsilon})$ \emph{independent} of the initial distance from the
homoclinic orbit.
The total error introduced by the averaging process in the first time-interval
is of
$O(\epsilon T_0)$ ($T_0$ is the period of the unperturbed orbit corresponding
to $\bar{E}(0)$,
the initial value of $\bar{E}$). A short explanation of this estimate is given
in section \ref{Argument: Averaging breaks down}.

For the second interval we can make use of the standard averaging theorems,
from which we get
that the introduced error on the second interval is of $O(\epsilon)$ and that
we are allowed
to extend the second interval to infinity, because all orbits are attracted to
the origin (see Sanders en Verhulst~\cite{S&V}, chapter 4). \\
This attracting property of the orbits also implies that the error introduced
from the first
does not blow up. Therefore, the total error introduced by the averaging
process is of
$O(\epsilon T_0)$ valid for all time.
As we will also show in section \ref{subsection:Approaching the boundary
layer}, for orbits
starting close to the homoclinic orbit of the unperturbed system $\bar{E}(0) =
\frac{1}{6}$,
we have that $T_0$ is of order $-\log(\frac{1}{6} - \bar{E}(0))$, which implies
that the averaged equation can be used to approximate the dynamics up to a
distance of
$O(\exp(-\frac{1}{\epsilon}))$ from the homoclinic orbit of the unperturbed
system.
More quantitative details on the boundary layer will be given
in section \ref{subsection:Approaching the boundary layer}.
We will also see in section \ref{subsection:Approaching the boundary layer}
that the averaged
equation indeed breaks down when we approach the boundary layer.

\subsubsection{Analysis of the averaged equation}

We now turn to the analysis of the averaged equation
(\ref{averaged:equation}).
The first
thing one should notice is that the effect of the decaying function $a(\epsilon
t)$ can be
removed from the equation by transforming to the new time $\tau$

\begin{equation}
  \label{averaged:transf:tau}
  \left\{
    \begin{aligned}
      \tau &= -\frac{1}{\epsilon} \log(a(\epsilon t)) \\
      a(\epsilon t) &= e^{-\epsilon \tau}
    \end{aligned}
  \right.
\end{equation}

Note that this transformation reduces to the identity transformation in the
special case
$a(\epsilon t) = e^{-\epsilon t}$. \\
It is remarkable that, given condition (\ref{restr:average:a}), it is not
important at all how $a(\xi)$ decays to zero, the dynamics of the
system does not change, apart from a rescaling of the time axis.

Applying this transformation produces the autonomous, 1-dimensional system

\begin{equation}
  \label{averaged:equation:tau}
  \frac{\mathrm{d}\bar{E}}{\mathrm{d}\tau} = -\epsilon A(\bar{E})
\end{equation}

We can solve this system explicitly by separation of variables, but
unfortunately
we do not have a primitive of $\frac{1}{A(\bar{E})}$ in the form of an
elementary function. \\
But we can draw some important conclusions from this system, of which the most
important
one is the existence of an \emph{adiabatic invariant}: As noted, it is
always possible to solve system (\ref{averaged:equation:tau}), which gives
the solution $\bar{E} = \bar{E}(\bar{E}(0), \epsilon \tau)$ as a function of
the initial
condition and slow time. Again, in principle one can solve this equation for
$\bar{E}(0)$
as a function of $\bar{E}$ and $\epsilon \tau$. Inverting the time
transformation
(\ref{averaged:transf:tau}), one finds $\bar{E}(0)$ as a function of $\bar{E}$
and
$\epsilon t$. Since $\bar{E}(0)$ is obviously time-independent, we reach the
conclusion that

\vspace{2ex}

\emph{There exists a global adiabatic invariant inside the homoclinic orbit of
the unperturbed system with the exclusion of an exponentially thin boundary
layer,
valid for all time, determined by equation (\ref{averaged:equation:tau}).}

\vspace{2ex}

\begin{figure}
  \begin{center}
    \includegraphics[height=7.5cm]{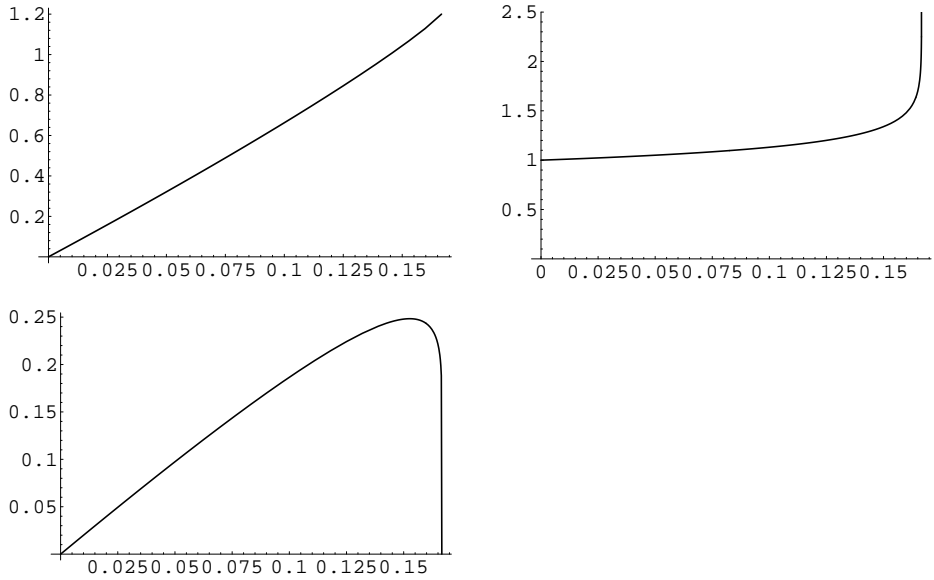}
  \end{center}
  \caption{The dependence of $\int{y_2}dy_1$, the
  $\frac{\mathrm{period}}{2\pi}$ and $A$ on $\bar{E}$}
  \label{Fig:dependence:E}
\end{figure}

For special cases we are able to produce these calculations explicitly, which
we will show now.
To understand these cases well, it is important to know how $\int{y_2}dy_1$,
the period and $A$
depend on $\bar{E}$. This is shown in figure \ref{Fig:dependence:E}.

It is clear that $\int{y_2}dy_1$ depends almost linearly on $\bar{E}$
throughout the entire
interval. This is understandable, since it is similar to the dependence of the
area of a disk on
its radius. What is not transparent is that the derivative of this function
goes to infinity as
$\bar{E}$ goes to $\frac{1}{6}$, but so slowly that its integral remains
bounded. \\
The period is close to $2\pi$ for small $\bar{E}$ as it should be, because in
this region the
unperturbed system behaves nearly like a harmonic oscillator with $\omega =
1$.
When $\bar{E}$ goes to $\frac{1}{6}$, the period goes to infinity, because the
orbits are approaching the saddle point, in the neighbourhood of which they
will stay a long time for
each passage. \\
The quotient of the two, $A$, shows the linear behaviour of $\int{y_2}dy_1$ for
small $\bar{E}$,
because the period is almost constant. However, $A$ has a maximum
($0.248320\ldots$) at $\bar{E} = 0.152640\ldots$, after which
it rapidly drops to zero. We could have predicted that $A$ is small for
$\bar{E}$ close to
$\frac{1}{6}$, since all the time the orbits are close to the saddle point, the
righthand side
of equation (\ref{timeder:integral:eps=0}) is small ($y_2 \ll 1$), resulting in
a small average.

\subsubsection{The adiabatic invariant}

We now turn to the calculation of the adiabatic invariant for $\bar{E}(0)$
small (we will make this
more precise later on). \\
To approximate the adiabatic invariant, we perform a Taylor expansion of
$A(\bar{E})$ around 0.
We note that the hypergeometric function forces us to use $\sqrt{\bar{E}}$
as expansion variable instead of just $\bar{E}$.
However, it turns out that the coefficients in front of the non-integer
powers of $\bar{E}$ are equal to zero, at least to fifth order. After a long
calculation we arrive at the following expansion, valid for $0 \leq \bar{E} <
\frac{1}{6}$

\begin{equation}
  A(\bar{E}) = 2 \bar{E} - \frac{5}{6} {\bar{E}}^2 - \frac{155}{54} {\bar{E}}^3
    - \frac{61135}{5184} {\bar{E}}^4 - \frac{825409}{15552} {\bar{E}}^5 +
    O({\bar{E}}^{5\frac{1}{2}})
\end{equation}

To approximate the adiabatic invariant, we truncate the series after the second
order terms, since
we are interested in the first non-trivial deviation from a slowly attracting
focus.
Substituting this quadratic expression into the averaged equation
(\ref{averaged:equation:tau}) we get

\begin{equation}
  \frac{\mathrm{d}\bar{E}}{\mathrm{d}\tau} = -2 \epsilon \bar{E} + \frac{5}{6}
  \epsilon {\bar{E}}^2
\end{equation}

which is easy to solve giving

\begin{equation}
  \bar{E}(\epsilon \tau) = \frac{2
  \bar{E}(0)}{\left(2-\frac{5}{6}\bar{E}(0)\right)e^{2 \epsilon \tau} +
  \frac{5}{6}\bar{E}(0)}
\end{equation}

 From this we readily obtain the adiabatic invariant

\begin{equation}
  \frac{\bar{E}(0)}{2 - \frac{5}{6}\bar{E}(0)} = \frac{\bar{E}(\epsilon
  \tau)}{2 - \frac{5}{6}\bar{E}(\epsilon \tau)} e^{2 \epsilon \tau}
\end{equation}

We are now able to specify what we meant with $\bar{E}(0)$ small. Since we have
neglected
$O({(\bar{E}(0))}^3)$ terms, we have introduced a new error of order
${(\bar{E}(0))}^3$ in the approximation of the solution.
Since we do not want this error term to dominate the error
introduced by the averaging process ($O(\epsilon)$), we take $\bar{E}(0)$ to be
$O({\epsilon}^{1/3})$.

Expanding the adiabatic invariant around $\bar{E} = 0$, we see that the first
non-trivial
correction to the slowly attracting focus (with adiabatic invariant $\bar{E}(0)
= \bar{E} e^{2\epsilon\tau}$) is given
by $\frac{5}{48}{\bar{E}}^2 e^{2\epsilon\tau}$ resulting in a slightly slower
collapse onto the
origin $(y_1, y_2) = (0, 0)$. \\
These arguments hold for the $(y_1, y_2)$ phase space only. To extend them to
the
original $(x_1, x_2)$ phase space, we have to invert the time-transformation
(\ref{averaged:transf:tau}) and the phase space transformation
(\ref{Trafo:x->y}), after
which we obtain the adiabatic invariant in the $(x_1, x_2)$ phase space:

\begin{equation}
  {{3 a(\epsilon t) {x_1^2} -
	2 {{a(\epsilon t)}^2} {x_1^3} +
	6 \epsilon a'(\epsilon t) x_1 x_2 +
	3 a(\epsilon t) {x_2^2} }\over
    {72 a(\epsilon t) - 15 {{a(\epsilon t)}^3} {x_1^2} +
      10 {{a(\epsilon t)}^4} {x_1^3} -
      30 \epsilon {{a(\epsilon t)}^2} a'(\epsilon t) x_1 x_2 -
      15 {{a(\epsilon t)}^3} {x_2^2}}}
\end{equation}

We include this rather lengthy expression, because it reveals an important
phenome\-non: Due to the
cross-terms $x_1 x_2$, the level curves of the adiabatic invariant
\emph{for a fixed time} ``resemble'' ellipses, of which the long axis and the
short axis differ by an $O(\epsilon)$ amount. and which are rotated around the
origin, causing asymmetry with respect to the $y_1$ and $y_2$ axis. \\
We did expect this for finite time, but this behaviour persists when we let $t$
tend to
infinity. Put in other words, when $t$ goes to infinity, our dynamical system
(\ref{Sys:x}) becomes symmetric (with respect to $x_1$ and $x_2$), but the
level curves of the adiabatic
invariant remain asymmetric. We have reached this important conclusion:

\vspace{2ex}

\emph{The evolution of an ensemble of phase points towards a symmetric
potential will show significant (i.e. $O(\epsilon)$) traces
of its asymmetrical past, for all time.}

\vspace{2ex}

So there is a sort of hysteresis effect present: although the
system becomes symmetric, it still ``knows'' that it was asymmetric in the
past. \\
We note that this phenomenon is not present in the $(y_1, y_2)$ phase space,
where the level curves of the adiabatic invariant are symmetric with respect to
the $y_1$-axis,
but is introduced by the phase space transformation (\ref{Trafo:x->y}) alone.

\begin{figure}
  \begin{center}
    \includegraphics[height=8cm]{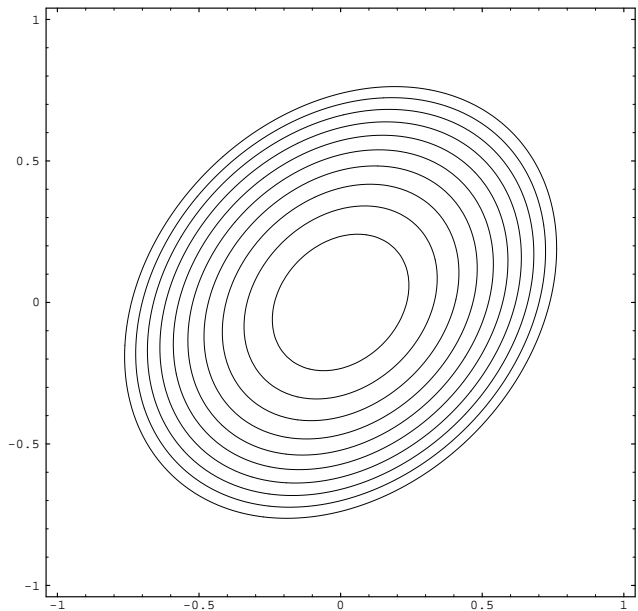}
  \end{center}
  \caption{A few level curves of the adiabatic invariant for $t$ fixed at
  infinity}
  \label{Fig:levelcurves:adiabaticinv}
\end{figure}

To demonstrate this phenomenon visually, we have to take $\epsilon$ not too
small, so we took
$\epsilon = \frac{1}{4}$. Figure \ref{Fig:levelcurves:adiabaticinv} shows a few
level curves
of the adiabatic invariant for $a(\epsilon t) = e^{-\epsilon t}$ and $t$ fixed
at infinity.
The asymmetric effect is clearly present.

As explained before, we expected to see effects of the slowly decaying
asymmetry in the
neighbourhood of the boundary layer separating the stable and unstable region,
but now it turns out that there are effects ($O(\epsilon)$) close to the origin
too.

\subsection{Approaching the boundary layer}
\label{subsection:Approaching the boundary layer}

We study the approach to the boundary layer, which is an $o(1)$ domain near
the homoclinic orbit and the stable manifold.

\begin{figure}
  \begin{center}
    \includegraphics[height=8cm]{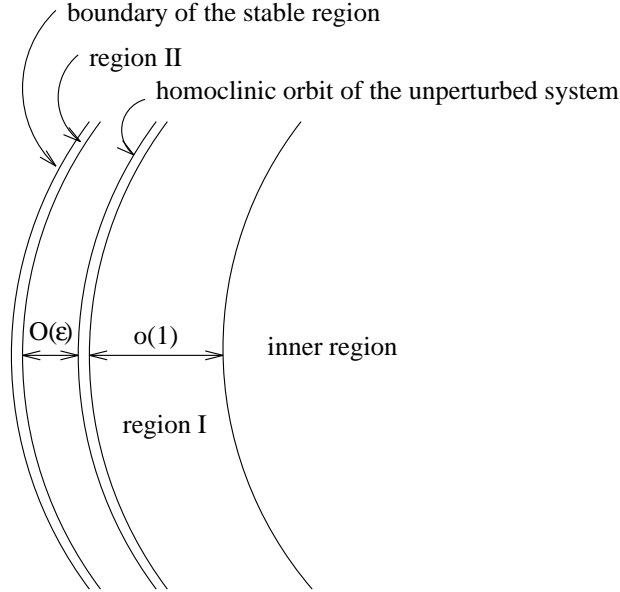}
  \end{center}
  \caption{The structure of the boundary layer}
  \label{Fig:structure boundary layer}
\end{figure}

More precisely, the boundary layer can be divided into three regions
(see figure \ref{Fig:structure boundary layer}). The first region consists of
the
phase points which are between $O(\exp(-\frac{1}{\epsilon}))$ and $o(1)$ inside
the homoclinic orbit of the unperturbed system.
It is in this region that the averaging technique slowly loses its validity, as
explained in
section \ref{Argument: Averaging breaks down}. We will call this region the
\emph{$o(1)$ boundary layer}.

The second region consists of the phase points which are within an
$O(\exp(-\frac{1}{\epsilon}))$ neighbourhood of
the boundary of the stable region. Orbits starting in these points will pass
the saddle point $(y_1, y_2) = (1, 0)$ on at least a $\frac{1}{\epsilon}$
timescale (which requires special attention), after which they will enter the
third region.
We will call this region the \emph{$O(\exp(-\frac{1}{\epsilon}))$ boundary
layer}.

The third region consists of the remaining phase points in the boundary layer,
which is a strip with an $O(\epsilon)$ width. Orbits starting in this region
will enter the first region
on an $O(1)$ timescale, which allows us to use the unperturbed orbits inside
this region.
We will call this region the \emph{$O(\epsilon)$ boundary layer}.

The \emph{inner region}, finally, consists of the phase points inside the
stable region but \emph{outside} the boundary layer.

Using the same approach as in the previous subsection, we are able to study the
adiabatic
invariant everywhere in the inner region and the $o(1)$ boundary layer. The
general idea is to expand the averaged equation
around a certain value of $\bar{E}$, in the neighbourhood of which we want the
study the
adiabatic invariant. This can be done to any desired precision. For low order
expansions,
it is possible to integrate the resulting equation explicitly. For high orders,
one has to use
numerical methods.

Approaching the boundary layer, there are two more special values of $\bar{E}$
which we will
study now, knowing the value of $\bar{E}$ corresponding to the maximum of
figure
(\ref{Fig:dependence:E}) and the maximum value $\bar{E} = \frac{1}{6}$.

The first special value can be computed numerically, giving ${\bar{E}}_{max} =
0.1526396\ldots$.
Expanding the averaged equation again to second order around ${\bar{E}}_{max}$,
we arrive at


\begin{equation}
  \frac{\mathrm{d}\bar{E}}{\mathrm{d}\tau} = -c_1 \epsilon + c_2 \epsilon
  {(\bar{E} - {\bar{E}}_{max})}^2
\end{equation}

with $c_1 = 0.2483204\ldots$ and $c_2 = 64.73966\ldots$, which is easy to solve
giving

\begin{equation}
  \bar{E} - {\bar{E}}_{max} = \sqrt{\frac{c_1}{c_2}} \tanh \left( \sqrt{c_1
  c_2} (-\epsilon \tau + I_{{\bar{E}}_{max}}) \right)
\end{equation}

where $I_{{\bar{E}}_{max}}$ is an integration constant determined by the
initial condition.
Solving $I_{{\bar{E}}_{max}}$ from this equation, we arrive again at the
adiabatic invariant:

\begin{equation}
  I_{{\bar{E}}_{max}} = \frac{1}{\sqrt{c_1 c_2}} \mathrm{artanh} \left(
  \sqrt{\frac{c_2}{c_1}} (\bar{E} - {\bar{E}}_{max}) \right) + \epsilon \tau
\end{equation}

Note that these equations hold only in an $O({\epsilon}^{1/3})$ neighbourhood
of ${\bar{E}}_{max}$, and therefore on an ${\epsilon}^{-2/3}$ timescale. For
instance, it does not make sense to take the limit $\tau \rightarrow \infty$.
Although this limit does exist, its value is obviously wrong.
Therefore the $\tanh$ should be regarded only as the first non-trivial
correction to the
linear time evolution of the adiabatic invariant around ${\bar{E}}_{max}$.

The second special value of $\bar{E}$, $\frac{1}{6}$, is much more interesting
and much more tricky, since it lies outside the domain of validity of the
averaging process. We can however still expand the averaged equation around
this value, because the part of the boundary layer inside the homoclinic orbit
of the
unperturbed system ($O(\exp(-\frac{1}{\epsilon}))$) is
much smaller than the domain of validity of the expansion
$O({\epsilon}^{1/2})$.
Therefore, we are allowed to use the results of this expansion, but only
\emph{outside} the boundary layer. Indeed we will see that the results of
this expansion inside the boundary layer are not correct.

Expanding the averaged equation around $\frac{1}{6}$ is not simple, because
the hypergeometric function has an infinite derivative at this point, and the
elliptic integral (the period) is unbounded at this point.

We break up this calculation by expanding equation (\ref{integral:y2:over:y1})
and equation (\ref{period:unperturbed:orbit}) separately.
After a straightforward calculation, we arrive at the following expansions:

\begin{equation}
  \label{expansions:E:1/6:int}
  \begin{aligned}
    \int{y_2}dy_1 = &\frac{6}{5} - 72 \left( 1 - \log \left( \frac{\frac{1}{6}
    - \bar{E}}{72} \right) \right) \left( \frac{\frac{1}{6} - \bar{E}}{72}
    \right) \\
    &+ O\left( \log \left( \frac{1}{6} - \bar{E} \right) {( \frac{1}{6} -
    \bar{E} )}^2 \right)
  \end{aligned}
\end{equation}

\begin{equation}
  \label{expansions:E:1/6:per}
  \begin{aligned}
    \mathrm{period} = &- \log \left( \frac{\frac{1}{6} - \bar{E}}{72} \right)
    -12 \left( 26 + 5 \log \left( \frac{\frac{1}{6} - \bar{E}}{72} \right)
    \right) \left( \frac{\frac{1}{6} - \bar{E}}{72} \right) \\
    &+ O\left( \log \left( \frac{1}{6} - \bar{E} \right) {( \frac{1}{6} -
    \bar{E} )}^2 \right)
  \end{aligned}
\end{equation}

Substituting these two expansions, we obtain for the averaged equation (with
$O\left( \epsilon {( \frac{1}{6} - \bar{E} )}^2 \right)$ terms neglected)

\begin{equation}
  \label{avg:equation:E:1/6}
  \frac{\mathrm{d}\bar{E}}{\mathrm{d}\tau} = \epsilon \frac{12}{5 \log \left(
  \frac{\frac{1}{6} - \bar{E}}{72} \right)} + 144 \epsilon \left( 1 -
  \frac{2}{\log \left( \frac{\frac{1}{6} - \bar{E}}{72} \right)} - \frac{26}{5
  \log^2 \left( \frac{\frac{1}{6} - \bar{E}}{72} \right)} \right) \left(
  \frac{\frac{1}{6} - \bar{E}}{72} \right)
\end{equation}

This equation is too complicated to be solved analytically. However, if we
neglect the $O\left( \epsilon ( \frac{1}{6} - \bar{E} ) \right)$ term too, it
is again possible to calculate the adiabatic invariant explicitly:

\begin{equation}
  \label{AI:E:1/6}
  I_{{\bar{E}}_{1/6}} = \left( \frac{\frac{1}{6} - \bar{E}}{72} \right) \log
  \left( \frac{\frac{1}{6} - \bar{E}}{72} \right) - \left( \frac{\frac{1}{6} -
  \bar{E}}{72} \right) + \frac{1}{30} \epsilon \tau
\end{equation}

Note that this adiabatic invariant is only valid on an
$\frac{1}{\sqrt{\epsilon}}$ timescale, since $\frac{1}{6} - \bar{E}$ will
become $O(\sqrt{\epsilon})$ on this timescale, causing an extra error of
$O(\epsilon)$.

At this point we are able to make some important remarks:

\begin{itemize}

\item{\emph{Every} orbit starting inside the $o(1)$ boundary layer will
collapse onto the attracting focus $(y_1, y_2) = (0, 0)$ on an
$\frac{1}{\epsilon}$ timescale, \emph{independent} of the initial distance from
the homoclinic orbit (collapsing onto the origin in the $(y_1, y_2)$ plane is
equivalent to circling around the origin in the $(x_1, x_2)$ plane). This
follows
directly from the adiabatic invariant (\ref{AI:E:1/6}), which forces the orbits
away from the boundary layer on an $\frac{1}{\epsilon}$ timescale.}

\item{The averaging process breaks down in the small strip between the $o(1)$
boundary layer and the homoclinic orbit of the unperturbed system, like we
expected it to. If the averaging process would be valid there too, \emph{every}
orbit starting there would collapse onto $(y_1, y_2) = (0, 0)$ on an
$\frac{1}{\epsilon}$ timescale, which would imply that all these orbits stay
within a certain bounded neighbourhood of the origin in the $(x_1, x_2)$
plane.
This cannot be true of course, because an orbit starting very close to the
saddle point $(x_1, x_2) = (1, 0)$ inside the homoclinic orbit, will end up
arbitrary far away from the origin in the $(x_1, x_2)$ plane.}

\item{The leading order behaviour of the period near the homoclinic orbit is
given by $- \log \left( \frac{\frac{1}{6} - \bar{E}}{72} \right)$. This is the
cause of the break-down of the averaging process, since averaging is only valid
if the period is $o(\frac{1}{\epsilon})$.}

\end{itemize}

\section{The boundary layer}
\label{Section: The boundary layer}

After the previous study of the major part of the stable region, we will turn
our attention to
the remaining part of the boundary layer. Since the $o(1)$ boundary layer is
covered by the previous
section, we only have to study the $O(\epsilon)$ and
$O(\exp(-\frac{1}{\epsilon}))$ boundary layers.
As we explained in the previous section, we cannot use the theory of averaging
for this study.

We treat the $O(\epsilon)$ and $O(\exp(-\frac{1}{\epsilon}))$ boundary layers
simultaneously. The
only difference between them is that orbits starting inside the
$O(\exp(-\frac{1}{\epsilon}))$
boundary layer will pass the the saddle point $(y_1, y_2) = (1, 0)$ on at least
a $\frac{1}{\epsilon}$
timescale, which results in an arbitrary large circle in the $(x_1, x_2)$ plane
as t tends to infinity. However this does
not require a separate treatment.

It is important to note that orbits starting inside the $O(\epsilon)$ boundary
layer will remain
within an $O(1)$ neighbourhood of the origin in the $(x_1, x_2)$ plane. So,
although the
$O(\epsilon)$ boundary layer appears to be larger than the
$O(\exp(-\frac{1}{\epsilon}))$ boundary
layer, it is in fact much smaller, because the latter has to fill up the rest
of the $(x_1, x_2)$ phase space.

To study the boundary layer, we can use the same method we used to compute the
position of the
boundary of the stable region, because the orbits in the boundary layer are
close to the
homoclinic orbit of the unperturbed problem. So, to calculate the orbits to
$O(\epsilon)$
precision, we are allowed to substitute expressions for the homoclinic orbit
into the $O(\epsilon)$
contributions to the dynamics.

This way we get again a two stage process. The first stage is governed by the
homoclinic orbit
of the unperturbed system. After the orbit has entered the domain of validity
of the averaging
process, the orbit collapses onto the origin on a $\frac{1}{\epsilon}$
timescale.

Since the existence of an adiabatic invariant was of great help in our study of
the inner region,
we prefer to extend that approach to the boundary layer. We expect an adiabatic
invariant to be
present in the boundary layer too, because we are studying a Hamiltonian system
which depends
adiabatically on time. Finding this adiabatic invariant is generally very hard
in regions where the
unperturbed system has non-periodic solutions (in our case, outside the
homoclinic orbit).

The straightforward way to find the adiabatic invariant is to perturb the
energy of the unperturbed
system in such a way that its time-derivative becomes $O(\epsilon^2)$. So we
are looking for an adiabatic invariant of the form

\begin{equation}
  \label{AI:bl}
  I_{bl}(y_1, y_2, \epsilon t) =  E(\epsilon = 0) + \epsilon g(E(\epsilon = 0),
  y_1, \epsilon t)
\end{equation}

where $E(\epsilon = 0)$ is given by (\ref{Def:integral:eps=0}).

By demanding that the time-derivative of $I_{bl}$ has a zero $O(\epsilon)$
contribution, one
normally arrives at a first order linear PDE for the function g. With a little
bit of foresight, we
choose the first argument to be orthogonal to the characteristic lines of the
PDE, which is why we
arrive at a first order linear ODE for the function g.

By using Gradshteyn and Ryzhik~\cite{G&R} intensively we derived this explicit
expression for the function g:

\begin{equation}
  \begin{aligned}
    g(E(\epsilon &= 0), y_1, \epsilon t) = - \frac{8}{15} \sqrt{\frac{2}{3}}
    \frac{a'(\epsilon t)}{a(\epsilon t)} \times \\
    &\times \left\{ \left( - b \xi + \frac{3}{2} \xi^3 \right)
    \sqrt{{(\xi^2-b)}^2+c^2} +
    (b^3 + b c^2) I_1 -\frac{9}{4} I_2 \right\} \\
    I_1 &= \frac{1}{2} {(b^2 + c^2)}^{-1/4} \sideset{_2}{_1}F(\alpha, r) \\
    I_2 &= \frac{1}{2} {(b^2 + c^2)}^{1/4} \left( \sideset{_2}{_1}F(\alpha, r)
    - 2 E(\alpha, r) \right)
    + \frac{\sqrt{{(\xi^2-b)}^2+c^2}}{\xi + \sqrt{b^2 + c^2} \xi^{-1}} \\
    \alpha &= \arccos \left\{ \frac{ \sqrt{b^2 + c^2} - \xi^2 }{ \sqrt{b^2 +
    c^2} + \xi^2 } \right\} \\
    r &= \frac{1}{2} \left( 1 + \frac{b}{\sqrt{b^2 + c^2}} \right)
  \end{aligned}
\end{equation}

Note that $E(\alpha, r)$ is the elliptic integral of the second kind and not
the energy. The real numbers $a$, $b$ and $c$ are related to the roots of a
third order polynomial in this way

\begin{equation}
  2 E(\epsilon = 0) - y_1^2 + \frac{2}{3} y_1^3 = \frac{2}{3} (y_1 - a)({(y_1 -
  a - b)}^2 + c^2) \qquad \forall y_1
\end{equation}

So $a$, $b$ and $c$ are functions of $E(\epsilon = 0)$, with $a < 0$, $b \geq
0$ and $c \geq 0$.
We want to make four remarks with respect to this formula:

\begin{itemize}
  \item We have derived the adiabatic invariant outside the homoclinic orbit
  (but inside the stable
	region) of the unperturbed
	system. It is however \emph{not} possible to calculate the adiabatic
	invariant in the
	inner region using the same procedure, since the characteristic lines
	are closed curves
	in the inner region, which prohibits the PDE to have a solution.
	Indeed, the adiabatic
	invariant we found previously for the inner region has an $O(\epsilon)$
	time-derivative.
  \item $I_{bl}$ determines the dynamics inside the boundary layer completely.
  This follows
	easily from $d(I_{bl}) = 0$.
  \item $I_{bl}$ is symmetric in $y_2$. Transforming back to the $(x_1, x_2)$
  plane introduces
	again the cross-terms $x_1 x_2$ in the adiabatic invariant which do not
	vanish for $t$
	going to infinity.
  \item We now have an adiabatic invariant throughout the entire stable region,
  with the exception
	of the very thin ($O(\exp(-\frac{1}{\epsilon}))$) region between the
	homoclinic orbit of the
	unperturbed system and the $o(1)$ boundary layer. This is not a
	problem, since we can
	approximate the dynamics inside this strip with transversal orbits
	which introduces only an
	$O(\frac{1}{\epsilon} \exp(-\frac{1}{\epsilon}))$ error. This trick
	solves the problem of matching the two adiabatic invariants at the same
	time.
\end{itemize}

We would like to visualize the dynamics going on inside the boundary layer.
Density
functions are not very useful for this, since our system is Hamiltonian which
implies area
conservation. This is well known for autonomous and time-periodic Hamiltonian
systems. To
prove it for general time-dependent systems, one introduces a new independent
variable equal
to the time (making the system autonomous). Applying Liouville's theorem proves
the desired
result.

Note that the conservation of area implies that the area of the ``tongue'' of
the boundary
layer is infinite, since it has to fill up the entire $(x_1, x_2)$ phase space
in the end.
So our ``thin'' boundary layer is in fact the largest part of the stable
region.

\begin{figure}
  \begin{center}
    \includegraphics[height=8cm]{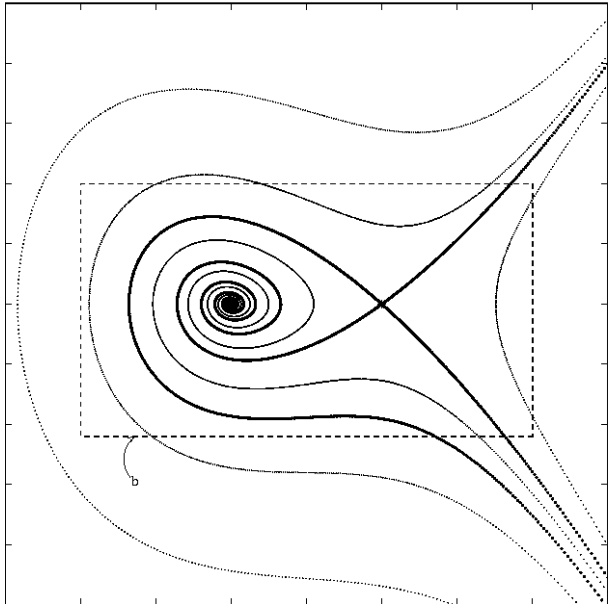}
  \end{center}
  \caption{The rectangular box around the homoclinic orbit of the unperturbed
  system.}
  \label{Fig:evolution:box}
\end{figure}

To study the dynamics inside the boundary layer, we therefore choose to look at
the evolution
of the rectangular box around the
homoclinic orbit of the unperturbed system, as depicted in figure
\ref{Fig:evolution:box}.

Since all orbits starting inside the box will remain inside the (evoluted) box,
we only have
to study the boundary of the box. Moreover, we only have to study those points
of the boundary
lying inside the stable region, since all other points clear off to infinity on
an $O(1)$
timescale. Therefore we only have to study the bottom boundary (b) of the box.

So by studying only a very limited set of phase space, we will gain information
about all
orbits starting inside the box, i.e. both inside the domain of validity of
averaging and
inside the boundary layer.

\begin{figure}
  \begin{center}
    \includegraphics[width=12.5cm]{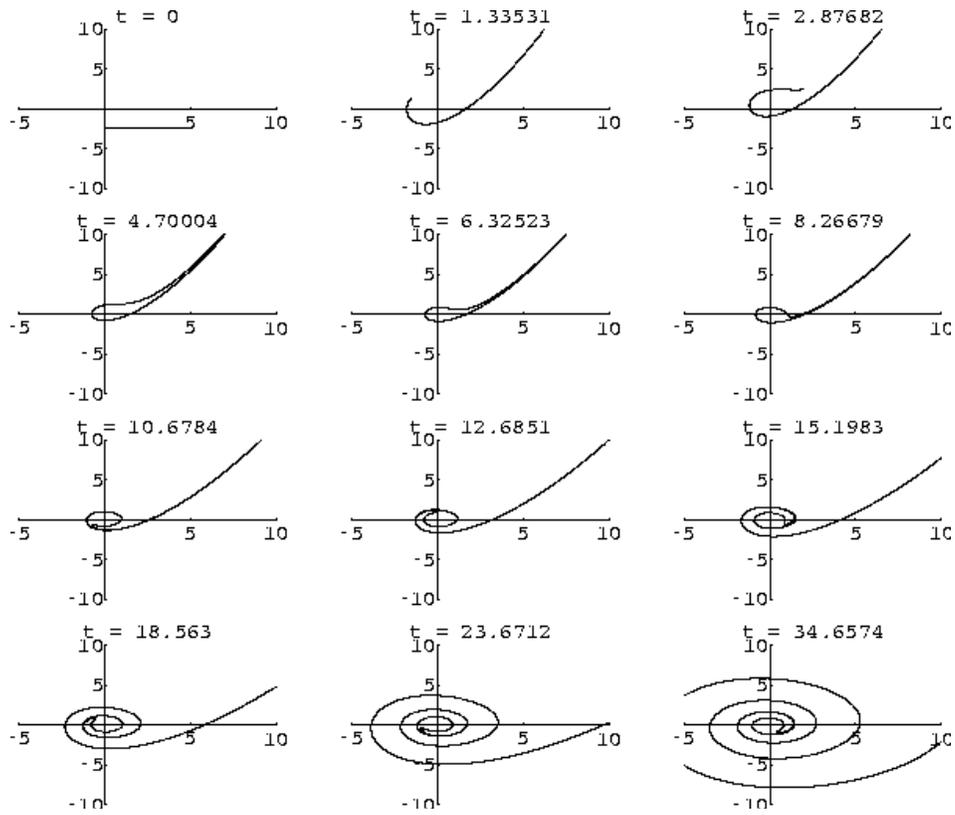}
  \end{center}
  \caption{The evolution of a straight line (part of the bottom boundary b)
  crossing the boundary of the stable region.}
  \label{Fig:evolution}
\end{figure}

For numerical reasons, we followed the evolution of the bottom boundary of a
different (but similar) box in the $(x_1, x_2)$ phase space, namely the
straight line between $(x_1, x_2) = (0, -2.5)$ and $(x_1, x_2) = (5, -2.5)$.
The numerical results are shown
in figure \ref{Fig:evolution}. We took $\epsilon = 0.1$ which forced us to take
steps along
the boundary as small as $10^{-14}$ to generate the last sub-figure. This is
due to the fact
that the most interesting dynamics takes place in an
$O(\exp(-\frac{1}{\epsilon}))$
neighbourhood of the boundary of the stable region.

The open area around the origin is the domain of validity of averaging. This is
the part of
phase space where the level curves of the adiabatic invariant (figure
\ref{Fig:levelcurves:adiabaticinv}) live.
It is also clear to see the instantaneous saddle point moving from $(1,0)$ to
infinity.

The points connecting the instantaneous saddle point with the domain of
validity of averaging
have started very close ($O(\exp(-\frac{1}{\epsilon}))$) to the boundary of the
stable region,
passed the saddle point during a time-interval of $O(\frac{1}{\epsilon})$,
after which they
entered the domain of validity of averaging (in the $(y_1, y_2)$ phase space).

The effect in the $(x_1, x_2)$ phase space is that the orbits end up circling
around the
origin outside the part where the level curves of the adiabatic invariant
(figure \ref{Fig:levelcurves:adiabaticinv}) live.
The closer an orbit starts to the boundary of the stable region, the larger the
radius of the
circle it describes in the end.

The effect of the area conservation is also nicely visible. Since the starting
box has a finite
area, the area inside the spiral must be finite too, which makes the spiral
very thin.
Note that from $t = 4.7$ on the curve going to infinity actually consists of
two very close curves.\\
Note also that the remaining (major) part of phase space has to be filled by
the tail of the
``small'' tongue of the boundary layer which lies outside the box.

\section{Concluding remarks}

It is surprising that it is possible to give a fairly complete treatment of
system (\ref{Sys:x}) which describes the evolution of a simple system with an
asymmetric potential to a symmetric potential. The most remarkable result is
that in the evolution towards symmetry as time tends to infinity, traces of the
asymmetric past can be recognized in the solutions.

System (\ref{Sys:x}) is just a metaphor for simplified models with two degrees
of freedom which exhibit evolution from asymmetry towards symmetry. In a
forthcoming paper we shall discuss such higher dimensional problems using
basically the same methods.

In the discussion of the validity of the averaged equation, we have assumed
that the reader is familiar with the proof of the standard averaging theorems
(see for instance Sanders and Verhulst~\cite{S&V}).
In particular, we have used the straightforward extension of those theorems to
periods depending on $\epsilon$: by rescaling the time-variable it is easily
shown that averaging produces $O(\epsilon T(\epsilon))$ approximations, valid
on a $\frac{1}{\epsilon}$ timescale, as long as $\epsilon T(\epsilon)$ is
$o(1)$.
\label{Argument: Averaging breaks down}

\newpage
\listoffigures

\end{document}